\documentclass[epj]{svjour}
\usepackage{graphics}
\begin{document}
\title{$CP$ violation for $B^{0}\rightarrow \rho^{0}(\omega)\rho^{0}(\omega) \rightarrow \pi^+\pi^-\pi^+\pi^-$ in QCD factorization}
\author{Gang L\"{u}\inst{1} % etc
\thanks{{e-mail:ganglv66@sina.com } }%
\and Jia-Qi Lei\inst{1}  % etc
\thanks{{e-mail:18236915407@139.com
} }%
\and Xin-Heng Guo\inst{2}   % etc
\thanks{{e-mail: xhguo@bnu.edu.cn } }%
\and Zhen-Hua Zhang\inst{3}  % etc
\thanks{{e-mail:zhangzh@iopp.ccnu.edu.cn} }%
\and Ke-Wei Wei\inst{3}  % etc
\thanks{{e-mail:weikw@hotmail.com} }%
}                     % Do not remove
%
%\offprints{}          % Insert a name or remove this line
%
\institute{College of Science, Henan University of Technology, Zhengzhou 450001, China
 \and College of Nuclear Science and Technology, Beijing
Normal University, Beijing 100875, China
\and School of Nuclear Science and Technology, University of South China,
 Hengyang, Hunan 421001, China
\and College of Physics and Electrical Engineering, Anyang Normal University, Anyang 455000, China }
\date{Received: date / Revised version: date}
% The correct dates will be entered by Springer
%
\abstract{In the QCD factorization (QCDF) approach we study the direct $CP$ violation in
$\bar{B}^{0}\rightarrow\rho^0(\omega)\rho^0(\omega)\rightarrow\pi^+\pi^-\pi^+\pi^-$
via the $\rho-\omega$ mixing mechanism. We find that the $CP$ violation can be
enhanced by double  $\rho-\omega$ mixing when the masses of the $\pi^+\pi^-$ pairs are in the vicinity of the $\omega$
resonance, and the maximum $CP$ violation can
reach 28{\%}. We also compare the results from the naive factorization
 and the QCD factorization.
\PACS{{11.30.Er}, {12.39.-x}, {13.20.He}, {12.15.Hh}
%      {PACS-key}{describing text of that key}   \and
%      {PACS-key}{describing text of that key}
     } % end of PACS codes
} %end of abstract
\maketitle

\section{\label{intro}Introduction}

$CP$ violation is an extensive research topics in recent years.
In Standard Model (SM), $CP$ violation is related to the
weak complex phase in the Cabibbo-Kobayashi-Maskawa (CKM) matrix
\cite{cab,kob}. In the past few years more attention has been focused on the decays of $B$
meson system both theoretically and experimentally.
Recently, the large $CP$ violation was found by the
LHCb Collaboration in the three-body
decay channels of $B^{\pm}\rightarrow \pi^{\pm}\pi^{+}\pi^{-}$
and $B^{\pm}\rightarrow K^{\pm}\pi^{+}\pi^{-}$\cite{J-R.A,J-R.A-1}.
Hence, the theoretical mechanism for the three or four-body decays
become more and more interesting. In this paper, we focus on the interference from
intermediate $\rho$ and $\omega$ mesons in the four-body decay.

It is known that the naive factorization \cite{bauer,bauer-1},
the QCD factorization (QCDF) \cite{qcdf,qcdf1,qcdf1-1},
the perturbative QCD (PQCD) \cite{pqcd,pqcd-1,pqcd1},
and the soft-collinear effective theory (SCET)\cite{scet,scet-1} are
the most extensive approaches for calculating
the hadronic matrix elements. These factorization approaches
present different methods for dealing with the hadronic matrix
 elements in the leading power of $1/m_b$ ($m_b$ is the b-quark mass).
Direct $CP$ violation occurs through the interference of two amplitudes with
different weak phases and strong phases. The weak phase difference
is directly determined by the CKM matrix elements, while the
strong phase is usually difficult to control.
However, and not well determined from a
theoretical approach. The B meson decay amplitude involves
the hadronic matrix elements which computation is not trivial.
Different methods may present different strong phases.
Meanwhile, we can also obtain a large strong phase difference
 by some phenomenological mechanism. $\rho-\omega$ mixing has been used for this purpose
in the past few years \cite{eno,gar,guo1,guo2,guo11,lei,gang1,gang2,gang3,gang4,gang5}.
In this paper, we will investigate the $CP$ violation via double $\rho-\omega$
mixing in the QCDF approach.

In the QCDF approach, at the rest frame of the heavy B meson, B meson can decay into two light
mesons with large momenta.
In the heavy-quark limit, QCD corrections can be calculated
for the non-leptonic two-body B-meson decays.  The decay amplitude
can be obtained at the next-to-leading power in $\alpha_{s}$ and the leading power in $\Lambda_{QCD}/m_{b}$.
In the QCD factorization, there is cancellation of the scale and renormalizaion scheme
dependence  between the Wilson coefficients and the hadronic matrix elements.
However, this does not happen in the naive factorization. The hadronic matrix elements can
be expressed in terms of form factors and meson light-cone distribution amplitudes
including strong interaction corrections.

The remainder of this paper is organized as follows. In Sec.
\ref{sec:hamckm} we present the form of the effective Hamiltonian.
In Sec. \ref{sec:cpv1} we give the calculating formalism of $CP$ violation from $\rho-\omega$ mixing
in $B^{0}\rightarrow\rho^0(\omega)\rho^0(\omega)\rightarrow\pi^+\pi^-\pi^+\pi^-$.
Input parameters are presented in Sec.\ref{input}.
We present the numerical rusults in Sec.\ref{sec:numerical}.
Summary and discussion are included in
Sec. \ref{sec:conclusion}.

\section{\label{sec:hamckm}The effective
hamiltonian}

With the operator product expansion, the effective weak Hamiltonian can
be written as \cite{Buch}
\begin{eqnarray}
 {\cal H}_{\rm eff}&=&\frac{G_{F}}{\sqrt{2}}
[\sum\limits_{p=u,c}\sum\limits_{q=d,s}V_{pb}V_{pq}^{*}(c_{1}O_{1}^{p}+c_{2}O_{2}^{p} \nonumber\\
&+& \sum\limits_{i=3}^{10}c_{i}O_{i}+c_{7\gamma}O_{7\gamma}+c_{8g}O_{8g}]
 +H_{.}c_{.},
\end{eqnarray}
where $G_F$ represents the Fermi constant, $c_{i}$ $(i=1,....,10$, $7\gamma,8g)$ are the Wilson coefficients, $V_{pb}$,
$V_{pq}$ are the CKM matrix elements. The
operators $O_i$ have the following forms:
%\begin{widetext}
\begin{equation}
\begin{array}{lll}
&&O_{1}^{p}=\bar{p}\gamma_{\mu}(1-\gamma_{5})b
\bar{q}\gamma^{\mu}(1-\gamma_{5})p,\\
&&O_{2}^{p}=\bar{p}_{\alpha}\gamma_{\mu}(1-\gamma_{5})b_{\beta}
\bar{q}_{\beta}\gamma^{\mu}(1-\gamma_{5})p_{\alpha}, \\
&&O_{3}=\bar{q}\gamma_{\mu}(1-\gamma_{5})b
\sum\limits_{q'}\bar{q'}\gamma^{\mu}(1-\gamma_{5})q',\\
&&O_{4}=\bar{q}_{\alpha}\gamma_{\mu}(1-\gamma_{5})b_{\beta}
\sum\limits_{q'}\bar{q'}_{\beta}\gamma^{\mu}(1-\gamma_{5})q'_{\alpha},\\
&&O_{5}=\bar{q}\gamma_{\mu}(1-\gamma_{5})b
\sum\limits_{q'}\bar{q'}\gamma^{\mu}(1+\gamma_{5})q',\\
&&O_{6}=\bar{q}_{\alpha}\gamma_{\mu}(1-\gamma_{5})b_{\beta}
\sum\limits_{q'}\bar{q'}_{\beta}\gamma^{\mu}(1+\gamma_{5})q'_{\alpha},\\
&&O_{7}=\frac{3}{2}\bar{q}\gamma_{\mu}(1-\gamma_{5})b
\sum\limits_{q'}e_{q'}\bar{q'}\gamma^{\mu}(1+\gamma_{5})q',\\
&&O_{8}=\frac{3}{2}\bar{q}_{\alpha}\gamma_{\mu}(1-\gamma_{5})b_{\beta}
\sum\limits_{q'}e_{q'}\bar{q'}_{\beta}\gamma^{\mu}(1+\gamma_{5})q'_{\alpha},\\
&&O_{9}=\frac{3}{2}\bar{q}\gamma_{\mu}(1-\gamma_{5})b
\sum\limits_{q'}e_{q'}\bar{q'}\gamma^{\mu}(1-\gamma_{5})q',\\
&&O_{10}=\frac{3}{2}\bar{q}_{\alpha}\gamma_{\mu}(1-\gamma_{5})b_{\beta}
\sum\limits_{q'}e_{q'}\bar{q'}_{\beta}\gamma^{\mu}(1-\gamma_{5})q'_{\alpha},  \\
&&O_{7\gamma}=\frac{-e}{8\pi^{2}}m_{b}\bar{s}\sigma_{\mu\nu}(1+\gamma_5)F^{\mu\nu}b,\\
&&O_{8g}=\frac{-g_s}{8\pi^{2}}m_{b}\bar{s}\sigma_{\mu\nu}(1+\gamma_5)G^{\mu\nu}b,
\end{array}
\end{equation}
%\end{widetext}
where $\alpha$ and $\beta$ are color indices, $O_{1}^{p}$ and
$O_{2}^{p}$ are the tree operators, $O_{3}-O_{6}$ are QCD penguin
operators which are isosinglets, $O_{7}-O_{10}$ arise from
electroweak penguin operators which have both isospin $0$ and $1$
components. $O_{7\gamma}$ and $O_{8g}$ are the electromagnetic and chromomagnetic
dipole operators, $e_{q'}$ are the electric charges of the quarks and $q'=u,d,s,c,b$
is implied.

For the decay channel $B^{0}\rightarrow\rho^0(\omega)\rho^0(\omega)$,
neglecting power corrections of order $\Lambda_{\rm QCD}/m_b$,
the transition
matrix element of an operator ${O}_i$ in the weak effective
Hamiltonian is given by {\cite{qcdf1,qcdf1-1}}
\begin{eqnarray}
\label{fff}
\langle V_1 V_2|{O}_i|\bar{B}\rangle &=&
\sum_j F_j^{B\to V_1}(m_{V_2}^2)\,\int_0^1 du\,T_{ij}^I(u)\,\Phi_{V_2}(u)\nonumber\\
\,\,&+&\,\,(V_1\leftrightarrow V_2)\nonumber\\
&&\hspace*{-2cm}
+\,\int_0^1 d\xi du dv \,T_i^{II}(\xi,u,v)\,
\Phi_B(\xi)\,\Phi_{V_1}(v)\,\Phi_{V_2}(u) \nonumber\\
\end{eqnarray}
Here $F_j^{B\to V_{1,2}}(m_{V_{2,1}}^2)$ denotes $B\to V_{1,2}$ ($V_{1,2}$ represent $\rho^0$ and $\omega$ mesons) form factor,
and $\Phi_V(u)$ is the light-cone distribution amplitude for the
quark-antiquark Fock state of mesons $\rho^0$ and $\omega$.
$T_{ij}^I(u)$ and $T_i^{II}(\xi,u,v)$ are hard-scattering functions,
which are perturbatively calculable. The hard-scattering kernels and
light-cone distribution amplitudes (LCDA) depend on the factorization scale
and the renormalization scheme. $m_{V_{1,2}}$ denote the $\rho^0$ and $\omega$ masses, respectively.

We match the effective weak Hamiltonian onto a transition operator, the matrix element is given by
($\lambda_p^{(D)}=V_{pb}V_{pD}^{*}$ with $D=d$) \cite{qcdf1,qcdf1-1}
\begin{equation}\label{Top}
   \langle V_1V_2|{\cal H}_{\rm eff}|\bar B\rangle
   = \sum_{p=u,c} \lambda_p^{(D)}\,
   \langle V_1 V_2|{\cal T}_A^{p,h} + {\cal T}_B^{p,h}|\bar B\rangle \,.
\end{equation}
where ${\cal T}_A^{p,h}$ denotes the contribution from vertex correction, penguin amplitude and spectator scattering in terms of
the operators $a_{i}^{p,h}$, ${\cal T}_B^{p,h}$ refers to annihilation terms contribution by operators $b_{i}^{p,h}$. $h$ is
the helicity of the final state.

The flavor operators $a_i^p$ are defined in \cite{qcdf1,qcdf1-1} as
follows:
\begin{eqnarray}
   a_i^{p,h}(V_1 V_2) &=& \left( c_i + \frac{c_{i\pm 1}}{N_c} \right)
   N_{i}^{h}(V_2) \nonumber\\
  && + \,\frac{c_{i\pm 1}}{N_c}\,\frac{C_F\alpha_s}{4\pi}
   \left[ V_{i}^{h}(V_2) + \frac{4\pi^2}{N_c}\,H_{i}^{h}(V_1 V_2) \right]\nonumber\\
   &&+ P_i^{p,h}(V_2) \,,
\end{eqnarray}
where $N_{c}$ is the number of colors, the upper (lower) signs
apply when $i$ is odd (even), and $C_{F}=\frac{N_c^2-1}{2N_c}$. It is
understood that the superscript `$p$' is to be omitted for $i=1,2$. The
quantities $V_i^h(V_2)$ account for one-loop vertex corrections,
$H_i^{h}(V_1 V_2)$ for hard spectator interactions, and $P_i^{p,h}(V_2)$ for
penguin contractions. $ N_i^h(V_2)$ is given by
\begin{equation}\label{loterms}
   N_i^h(V_2) = \Bigg\{
   \begin{array}{ll}
    ~0 \,; & \quad \mbox{$i=6,8 $,} \\
    ~1 \,; & \quad \mbox{all other cases.}
   \end{array}
\end{equation}

The coefficients of the flavor operators $\alpha_i^{p,h}$ can be expressed in terms of the
coefficients $a_i^{p,h}$. We will present the form in the following section. Using the unitarity relation
\begin{eqnarray}
\lambda_u^{(D)}+\lambda_c^{(D)}+\lambda_t^{(D)}=0,
\end{eqnarray}
we can get
%\begin{widetext}
\begin{eqnarray}\label{alphaidef4}
&&\sum_{p=u,c}\lambda_p^{(D)} {\cal T}_A^{p,h} \nonumber\\
 &=&\sum_{p=u,c} \lambda_p^{(D)}\Bigg[\delta_{pu}\,\alpha_1(V_1 V_2)\,A([\bar q_s u][\bar u D]) \nonumber\\
    &+&\delta_{pu}\,\alpha_2(V_1 V_2)\,A([\bar q_s D][\bar u u])\Bigg]
     +  \lambda_u^{(D)} \Bigg[(\alpha_4^u(V_1 V_2) \nonumber\\
     &-&\alpha_4^c(V_1 V_2))\,\sum_q A([\bar q_s q][\bar q D])
   +(\alpha_{4,\rm EW}^u(V_1 V_2) \nonumber\\
   &-&\alpha_{4,\rm EW}^c(V_1 V_2))\,\sum_q\frac32\,e_q\,
    A([\bar q_s q][\bar q D]) \Bigg]  \nonumber\\
    &-& \lambda_t^{(D)}\Bigg[\alpha_3^c(V_1 V_2)\,\sum_q A([\bar q_s D][\bar q q]) \nonumber\\
    &+&\alpha_4^c(V_1 V_2)\,\sum_q A([\bar q_s q][\bar q D]) \nonumber\\
   &+&\alpha_{3,\rm EW}^c(V_1 V_2)\,\sum_q\frac32\,e_q\,
    A([\bar q_s D][\bar q q])   \nonumber\\
    &+& \alpha_{4,\rm EW}^c(V_1 V_2)\,\sum_q\frac32\,e_q\,
    A([\bar q_s q][\bar q D]) \Bigg],
\end{eqnarray}
%\end{widetext}
where the sums extend over $q=u,d,s$, and $\bar q_s$ denotes the
spectator antiquark.

Next we need to change the annihilation part into the following form  \cite{qcdf1,qcdf1-1}:

\begin{eqnarray}\label{bis}
   \sum_{p=u,c}\lambda_p^{(D)}{\cal T}_B^{p,h}
   &=& \sum_{p=u,c}\lambda_p^{(D)}    \nonumber\\
  &\times& \Bigg[\delta_{pu}\,b_1(V_1 V_2)\,\sum_{q'}
    B([\bar u q'][\bar q' u][\bar D b]) \nonumber\\
    &+& \delta_{pu}\,b_2(V_1 V_2)\,\sum_{q'}
    B([\bar u q'][\bar q' D][\bar u b]))\Bigg] \nonumber\\
  &-&\lambda_t^{(D)}\Bigg[ b_3(V_1 V_2)\,\sum_{q,q'} B([\bar q q'][\bar q' D][\bar q b])\nonumber\\
    &+& b_4(V_1 V_2)\,\sum_{q,q'} B([\bar q q'][\bar q' q][\bar D b])
    \nonumber\\
   &+& b_{3,\rm EW}(V_1 V_2)\,\sum_{q,q'} \frac{3}{2}\,e_q\,
    B([\bar q q'][\bar q' D][\bar q b])  \nonumber\\
    &+& b_{4,\rm EW}(V_1 V_2)\,\sum_{q,q'} \frac{3}{2}\,e_q\,
    B([\bar q q'][\bar q' q][\bar D b])\Bigg], \nonumber\\
\end{eqnarray}
where $b_{i}^{p,h}$, $b_{i,\rm EW}^{p,h}$ and $B$ will be given in the following section.

\section{\label{sec:cpv1}CP violation in $B^{0}\rightarrow\rho^0(\omega)\rho^0(\omega)\rightarrow\pi^+\pi^-\pi^+\pi^-$}
\subsection{Formalism}
The $B\rightarrow V_1(\epsilon_1,P_1)V_2(\epsilon_2,P_2)$ ($\epsilon_1$($P_1$) and $\epsilon_2$($P_2$) are the
polarization vectors (momenta) of $V_{1}$ and $V_{2}$, respectively) decay rate is
written as

\begin{eqnarray}
\Gamma=\frac{G_{F}^{2}P_{c}}{64\pi m_{B}^{2}}\sum_{\sigma}A^{(\sigma)+}A^{(\sigma)},
 \end{eqnarray}
where $P_{c}$ refers to the c.m. momentum. $A^{(\sigma)}$ is the helicity amplitude for each helicity of the final state.
The decay amplitude, $A$, can be decomposed into three components $H_{0}$, $H_{+}$, $H_{-}$ according to the helicity of the final state.
With the helicity summation, we can get
\begin{eqnarray}
\sum_{\sigma}A^{(\sigma)+}A^{(\sigma)}=|H_{0}|^{2}+|H_{+}|^{2}+|H_{-}|^{2}.
 \end{eqnarray}

In the vector meson dominance model
{\cite{Sakurai1969}}, the photon propagator is dressed by coupling
to vector mesons. Based on the same mechanism, $\rho-\omega$ mixing
was proposed {\cite{Connell1997,Connell1997-1}}. The formalism for $CP$ violation
in the decay of a bottom hadron, $\bar{B}$, will be reviewed in the
following. The amplitude for $\bar{B}\rightarrow V\pi^{+}\pi^{-}$,
$A$, can be written as
\begin{eqnarray}
A=\langle\pi^{+}\pi^{-}V|H^{T}|\bar{B}\rangle+\langle\pi^{+}\pi^{-}V|H^{P}|\bar{B}\rangle,
\end{eqnarray}
where $H^{T}$ and $H^{P}$ are the Hamiltonians for the tree and
penguin operators, respectively. We define the relative magnitude
and phases between these two contributions as follows:
\begin{eqnarray}
A=\langle\pi^{+}\pi^{-}V|H^{T}|\bar{B}\rangle[1+re^{i\delta}e^{i\phi}],
\end{eqnarray}
where $\delta$ and $\phi$ are strong and weak phase differences,
respectively. The weak phase difference $\phi$ arises from the
appropriate combination of the CKM matrix elements:
$\phi=\arg[(V_{tb}V_{td}^{*})/(V_{ub}V_{ud}^{*})]$.
The parameter $r$ is the absolute value of the ratio of tree and penguin
amplitudes,
\begin{eqnarray}
r=\left|\frac{\langle\pi^{+}\pi^{-}V|H^{P}|\bar{B}\rangle}{\langle\pi^{+}\pi^{-}V|H^{T}|\bar{B}\rangle}\right|.
\end{eqnarray}
The amplitude for $B\rightarrow\bar{V}\pi^{+}\pi^{-}$ is
\begin{eqnarray}
\bar{A}=\langle\pi^{+}\pi^{-}\bar{V}|H^{T}|B\rangle+\langle\pi^{+}\pi^{-}\bar{V}|H^{P}|B\rangle.
\end{eqnarray}
Then, the CP violating asymmetry, $A_{CP}$, can be written as
\begin{eqnarray}
A_{CP}&=&\frac{|A|^{2}-|\bar{A}|^{2}}{|A|^{2}+|\bar{A}|^{2}} \nonumber \\
  &=&\frac{-2(T_{0}^2r_{0}\sin\delta_0+T_{+}^2r_{+}\sin\delta_{+}+T_{-}^2r_{-}\sin\delta_-)\sin\phi}
  {\sum_{i=0+-}T_{i}^2(1+r_{i}^2+2r_{i}\cos\delta_i\cos\phi)},    \nonumber \\
\label{eq:CP-tuidao}
\end{eqnarray}
where
\begin{eqnarray}
|A|^{2}=\sum_{\sigma}A^{(\sigma)+}A^{(\sigma)}=|H_{0}|^{2}+|H_{+}|^{2}+|H_{-}|^{2}
\end{eqnarray}
and $T_{i}(i=0,+,-)$ represent the tree-level helicity amplitudes.
We can see explicitly from Eq. (\ref{eq:CP-tuidao}) that both weak and strong phase
differences are needed to produce $CP$ violation. $\rho-\omega$
mixing has the dual advantages that the strong phase difference is
large and well known {\cite{eno,gar}}. In this
scenario one has
\begin{eqnarray}
\langle\pi^{+}\pi^{-}\pi^{+}\pi^{-}|H^{T}|\bar{B}\rangle=\frac{2g_{\rho}^{2}}{s_{\rho}^{2}s_{\omega}}\tilde{\Pi}_{\rho\omega}(t_{\omega}
+t_{\omega}^{a})\nonumber\\
+\frac{g_{\rho}^{2}}{s_{\rho}^{2}}(t_{\rho}+t_{\rho}^{a}),
\label{tree-scena1}
\end{eqnarray}
\begin{eqnarray}
\langle\pi^{+}\pi^{-}\pi^{+}\pi^{-}|H^{P}|\bar{B}\rangle=\frac{2g_{\rho}^{2}}{s_{\rho}^{2}s_{\omega}}\tilde{\Pi}_{\rho\omega}(p_{\omega}
+p_{\omega}^{a})\nonumber\\
+\frac{g_{\rho}^{2}}{s_{\rho}^{2}}(p_{\rho}+p_{\rho}^{a}),
\label{pengui-scena1}
\end{eqnarray}
where $t_{V}(V=\rho$ or $ \omega)$ is the tree amplitude and $p_{V}$
is the penguin amplitude for producing a vector meson, $V$.
$t_{V}^{a}(V=\rho$ or $ \omega)$ is the tree annihilation amplitude and $p_{V}^{a}$
is the penguin annihilation amplitude. $g_{\rho}$ is the coupling for $\rho^{0}\rightarrow \pi^{+}\pi^{-}$,
$\tilde{\Pi}_{\rho\omega}$ is the effective $\rho-\omega$ mixing
amplitude, and $s_{V} $ is from the inverse propagator of the vector
meson V,
\begin{eqnarray}
s_{V}=s-m_{V}^{2}+i m_{V}\Gamma_{V},
\end{eqnarray}
with $\sqrt{s}$ being the invariant mass of the $\pi^{+}\pi^{-}$
pair.
The direct $\omega \rightarrow \pi^{+}\pi^{-}$ is effectively
absorbed into $\tilde{\Pi}_{\rho\omega}$, leading to the explicit
$s$ dependence of $\tilde{\Pi}_{\rho\omega}$ {\cite{Maltman1996,Maltman1996-1}}.
Making the expansion
$\tilde{\Pi}_{\rho\omega}(s)=\tilde{\Pi}_{\rho\omega}(m_{\omega}^{2})
+(s-m_{\omega}^{2}) \tilde{\Pi}'_{\rho\omega}(m_{\omega}^{2})$, the
$\rho-\omega$ mixing parameters were determined in the fit of
Gardner and O'Connell {\cite{S.Gardner1998}}: $\rm
Re\tilde{\Pi}_{\rho\omega}(m_{\omega}^{2})=-3500\pm300$ MeV$^{2}$,
$\rm Im\tilde{\Pi}_{\rho\omega}(m_{\omega}^{2})=-300\pm300$
MeV$^{2}$, and $\tilde{\Pi}'_{\rho\omega}(m_{\omega}^{2})=0.03\pm
0.04$. In practice, the effect of the derivative term is negligible.
From Eqs. (\ref{eq:CP-tuidao})(\ref{tree-scena1}), one has
\begin{eqnarray}
re^{i\delta}e^{i\phi}=\frac{2\tilde{\Pi}_{\omega}(p_{\omega}+p_{\omega}^{a})+s_{\omega}(p_{\rho}+p_{\rho}^{a})}
{2\tilde{\Pi}_{\rho\omega}(t_{\omega}+t_{\omega}^{a})+s_{\omega}(t_{\rho}+t_{\rho}^{a})}.
\label{jianhua1}
\end{eqnarray}
Defining
\begin{eqnarray}
\frac{t_{\omega}+t_{\omega}^{a}}{t_{\rho}+t_{\rho}^{a}}&=&\alpha e^{i\delta_{\alpha}}, \label{tree-form}\\
\frac{p_{\rho}+p_{\rho}^{a}}{p_{\omega}+p_{\omega}^{a}}&=&\beta e^{i\delta_{\beta}},\label{p-form} \\
\frac{p_{\omega}+p_{\omega}^{a}}{t_{\rho}+t_{\rho}^{a}}&=&r'e^{i(\delta_{q}+\phi)},\label{p2-form}
\end{eqnarray}
where $\delta_{\alpha}$, $\delta_{\beta}$, and $\delta_{q}$ are
strong phases, one finds the following expression from Eq. (\ref{jianhua1}):
\begin{eqnarray}
re^{i\delta}=r'e^{i\delta_{q}}\frac{2\tilde{\Pi}_{\rho\omega}+\beta
e^{i\delta^{\beta}}s_{\omega}
}{s_{\omega}+2\tilde{\Pi}_{\rho\omega}\alpha e^{i\delta_{\alpha}} }.
\label{jianhua2}
\end{eqnarray}
$\alpha e^{i\delta_{\alpha}}$, $\beta e^{i\delta_{\beta}}$, and
$re^{i\delta}$ will be calculated in the QCD factorization approach
in the next section. With Eq. (\ref{jianhua2}), we can obtain $r\sin\delta$ and $r\cos\delta$.
In order to get the $CP$ violating asymmetry, $A_{CP}$, in Eq.  (\ref{eq:CP-tuidao}),
$\sin\phi$ and $\cos\phi$ are needed. $\phi$ is determined by the
CKM matrix elements. In the Wolfenstein parametrization
{\cite{Wolfenstein1983,Wolfenstein1983-1}}, one has
\begin{eqnarray}
\sin\phi=\frac{\eta}{\sqrt{[\rho(1-\rho)-\eta^{2}]^{2}+\eta^{2}}},\\
\cos\phi=\frac{\rho(1-\rho)-\eta^{2}}{\sqrt{[\rho(1-\rho)-\eta^{2}]^{2}+\eta^{2}}}.
\end{eqnarray}.

\subsection{\label{sec:calculation}The calculation details}

In the QCD factorization approach, $\alpha_i$ associated with the coefficients $a_i$ can be written as follows
 (helicity indices are neglected){\cite{qcdf1,qcdf1-1}}:
\begin{eqnarray}
  \label{eq:alphaa}
    \alpha_1= a_1 \\
    \alpha_2= a_2 \\
    \alpha^p_3= a^p_3+a^p_5 \\
    \alpha^p_{3,EW}= a^p_9 + a^p_7  \\
    \alpha^p_4 = a^p_4- r_\chi^{V_2} a^p_6 \\
    \alpha^p_{4,EW}  = a^p_{10} - r_\chi^{V_2} a^p_8,
\end{eqnarray}
where we have used the notation
\begin{equation}
\label{rchidef}
  r_\chi^V \equiv \frac{2m_V}{m_b} \frac{f_V^\perp}{f_V}.
\end{equation}
with $f_V^\perp$, $f_V$ referring to the transverse decay constant and decay constant of the vector meson, respectively.

The flavor operators $a_i^{p,h}$
include short-distance nonfactorizable corrections such
as vertex corrections and hard spectator interactions.
 $V_2$ is the emitted meson
and $V_1$ shares the same spectator quark with the $B$ meson.

 The vertex corrections are given by{\cite{qcdf1,qcdf1-1}}:
\begin{equation}\label{vertex0}
   V^0_i(V_{2}) = \left\{\,\,
   \begin{array}{ll}
    {\displaystyle \int_0^1\!dx\,\Phi_\parallel^{V_2}(x)\,
     \Big[ 12\ln\frac{m_b}{\mu} - 18 + g(x) \Big]} \,, & \qquad \\ \\
     (i=\mbox{1--4},9,10) \\[0.4cm]
   {\displaystyle \int_0^1\!dx\,\Phi_\parallel^{V_2}(x)\,
     \Big[ - 12\ln\frac{m_b}{\mu} + 6 - g(1-x) \Big]} \,, & \qquad \\ \\
     (i=5,7) \\[0.4cm]
   {\displaystyle \int_0^1\!dx\, \Phi_{v_2}(x)\,\Big[ -6 + h(x) \Big]}
    \,, & \qquad \\ \\
    (i=6,8)
   \end{array}\right.
\end{equation}
\begin{equation}\label{vertexpm}
   V^\pm_i(V_2) = \left\{\,\,
   \begin{array}{ll}
    {\displaystyle \int_0^1\!dx\, \Phi^{V_2}_\pm(x) \,
     \Big[ 12\ln\frac{m_b}{\mu} - 18 + g_T(x) \Big]} \,, & \\
    \\\\ (i=\mbox{1--4},9,10) \\[0.4cm]
   {\displaystyle \int_0^1\!dx\, \Phi^{V_2}_\mp(x) \,
     \Big[ - 12\ln\frac{m_b}{\mu} + 6 - g_T(1-x) \Big]}\,, & \
     \\\\(i=5,7) \\[0.4cm]
  0,  (i=6,8)
   \end{array}\right.
\end{equation}
with
 \begin{eqnarray}
 g(x)&=& 3\Bigg( \frac{1-2x}{1-x} \ln x -i\pi\Bigg)\nonumber\\
  && + \Big[ 2{\rm Li}_2(x) -\ln^2x +\frac{2\ln x}{1-x}\nonumber\\
  && -(3+2i\pi)\ln x -
 (x \leftrightarrow 1-x)\Big]\,, \nonumber\\
 h(x)&=&  2{\rm Li}_2(x) -\ln^2x -(1+2i\pi)\ln x -
 (x \leftrightarrow 1-x)\,, \nonumber\\
 g_T(x)&=& g(x) + \frac{\ln x}{\bar x}\,,
 \end{eqnarray}
where $\bar x=1-x$, $\Phi^V_\parallel$ is a twist-2 light-cone distribution
amplitude of the meson $V$, $\Phi_{V_2}$ (for the longitudinal
component) and $\Phi_\pm$ (for transverse components) are twist-3
ones.

\vskip 0.2in \noindent{\it \underline{Hard spectator terms}} \vskip 0.1in

$H^h_i(V_1 V_2)$ arise from  hard spectator interactions with a hard
gluon exchange between the emitted  meson and the spectator quark of
the $\overline B$ meson. $H^0_i(V_1 V_2)$ have the expressions{\cite{qcdf1,qcdf1-1}}:
\begin{eqnarray}\label{eq:sepc01}
  H^0_i(V_1 V_2)&=& {if_B f_{V_1} f_{V_2} \over X^{(\overline{B} V_1,
  V_2)}_0}\,{m_B\over\lambda_B} \int^1_0 d u d v \,  \nonumber \\
 &&\Bigg( \frac{\Phi^{V_1}_\parallel(u) \Phi^{V_2}_\parallel(v)}{\bar u \bar v}
 + r_\chi^{V_1}
  \frac{\Phi_{v_1} (u) \Phi^{V_2}_\parallel(v)}{\bar u v}\Bigg),
 \hspace{0.5cm}
 \end{eqnarray}
for $i=1-4,9,10$,
\begin{eqnarray}\label{eq:spec02}
  H^0_i(V_1 V_2)&=& -{if_B f_{V_1} f_{V_2} \over X^{(\overline{B} V_1, V_2)}_0}
  \,{m_B\over\lambda_B}\int^1_0 d u d v \,       \nonumber \\
&& \Bigg( \frac{\Phi^{V_1}_\parallel(u) \Phi^{V_2}_\parallel(v)}{\bar u  v}
 + r_\chi^{V_1}
  \frac{\Phi_{v_1} (u) \Phi^{V_2}_\parallel(v)}{\bar u \bar v}\Bigg),
 \hspace{0.5cm}
 \end{eqnarray}
for $i=5,7$, and $H^0_i(V_1 V_2)=0$ for $i=6,8$.
 The transverse hard spectator terms $H^\pm_i(V_1
V_2)$ read
 \begin{eqnarray}
 H^-_i(V_1 V_2) &=& {2i f_B
 f^{\perp}_{V_1} f_{V_2} m_{V_2} \over m_B X^{({\overline B} {V_1}, V_2)}_-}
 \,{m_B\over\lambda_B}\int^1_0 dudv\,
 {\Phi^{V_{1}}_\perp(u)\Phi_-^{V_2}(v)\over \bar u^2 v}, \label{eq:H1m}  \nonumber \\
 H^+_i(V_1 V_2) &=& -\frac{2i f_B f_{V_1} f_{V_2} m_{V_1} m_{V_2}}{m_B^2
  X^{({\overline B} V_1, V_2)}_+}\,{m_B\over\lambda_B}  \nonumber \\
  &&\int^1_0 dudv\, {(\bar u-v)
  \Phi_+^{V_1}(u)\Phi_+^{V_2}(v)\over \bar u^2\bar v^2},   \nonumber\\
   \label{eq:H1p}
  \end{eqnarray}
 for  $i=1-4,9,10$,\\
 \begin{eqnarray}
 H^-_i(V_1 V_2) &=&-{2i f_B
 f^{\perp}_{V_1} f_{V_2} m_{V_2} \over m_B X^{({\overline B} {V_1}, V_2)}_-}
 \,{m_B\over\lambda_B}             \nonumber \\
&& \int^1_0 dudv\,
 {\Phi^{V_{1}}_\perp(u)\Phi_+^{V_2}(v)\over \bar u^2\bar v},\label{eq:H5m}
  \\
 H^+_i(V_1 V_2) &=& -\frac{2i f_B f_{V_1} f_{V_2} m_{V_1} m_{V_2}}{m_B^2
  X^{({\overline B} V_1, V_2)}_+}\,{m_B\over\lambda_B}   \nonumber \\
 && \int^1_0 dudv\, {(u-v)
  \Phi_+^{V_1}(u)\Phi_-^{V_2}(v)\over \bar u^2 v^2}, \label{eq:H5p}
  \end{eqnarray}
for $i=5,7$, and \\
 \begin{eqnarray} \label{eq:H6}
 H^-_i(V_1 V_2) &=&-{i f_B
 f_{V_1} f_{V_2} m_{V_2} \over m_B X^{({\overline B} {V_1}, V_2)}_-}
 {m_Bm_{V_1}\over m_{V_2}^2}\,
 \,{m_B\over\lambda_B}     \nonumber \\
 &&\int^1_0 dudv\,
 {\Phi^{V_{1}}_+(u)\Phi_\perp^{V_2}(v)\over v\bar u\bar v}, \label{eq:H6m} \\
 H^+_i(V_1 V_2) &=& 0,
  \end{eqnarray}
for $i=6,8$.
One can find that the expressions for
$H_i^{\pm}(V_1V_2)$ are independent of the choice for transverse
polarization vectors.

The helicity dependent factorizable amplitudes defined by
 \begin{eqnarray} \label{eq:Xamplitude}
 X^{(\bar{B}V_1,V_2)}=\langle V_2(p_2,\epsilon_2^*)|J_\mu|0\rangle\langle V_1(p_1,\epsilon_1^*)|J^\mu|B\rangle
\end{eqnarray}
have the expressions
\begin{eqnarray} \label{eq:Xh}
 X_0^{(\bar{B}V_1,V_2)}&=& \frac{if_{V_2}}{2m_{V_1}}[
 (m_{B}^2-m_{V_1}^2-m_{V_2}^2)(m_{B}+m_{V_{1}})\nonumber \\
 &&A_{1}^{BV_1}(q^2)
 -\frac{4m_{B}^2P_{c}^2}{m_{B}+m_{V_1}}A_{2}^{BV_1}(q^2)],
 \end{eqnarray}
\begin{eqnarray}
X_\pm^{(\bar{B}V_1,V_2)} &=& -if_{V_2}m_{B}m_{V_2}\Bigg[
\Bigg(1+\frac{m_{V_1}}{m_B}\Bigg)A_1^{BV_1}(q^2)   \nonumber \\
&&\mp\frac{2P_c}{
 m_{B}+m_{V_1}}V^{BV_1}(q^2)\Bigg],
 \end{eqnarray}
where $A_{i}^{BV_1}(i=1,2)$ and $V^{BV_1}$ are weak form factors.

\vskip 0.2in \noindent{\it \underline{Penguin terms}} \vskip 0.1in

At order $\alpha_s$, corrections from penguin contractions are
present only for $i=4,6$. For $i=4$ we have {\cite{qcdf1,qcdf1-1}}
\begin{eqnarray}\label{eq:PK}
   P_4^{h,p}(V_2) &=& \frac{C_F\alpha_s}{4\pi N_c}\Bigg\{
    c_1 [ G^h_{V_2}(s_p)+g_{V_2}]  \nonumber \\
   & &+ c_3 [  G^{h}_{V_2}(s_s) + G^h_{V_2}(1) +2g_{V_2}] \nonumber \\
    && + (c_4+c_6)\sum_{i=u}^b  \left[G^h_{V_2}(s_i)+g'_{V_2}\right]\nonumber \\
    && -2 c_{8g}^{\rm eff} G^h_g \Bigg \}
\end{eqnarray}
where $s_i=m_i^2/m_b^2$ and the function $G^h_{V_2}(s)$ is given by
\begin{eqnarray}  \label{eq:GK}
 G^h_{V_2}(s) &=&
 4\int^1_0 du\,\Phi^{V_2,h}(u)\int^1_0 dx\,x\bar x
  \ln[s-\bar u x\bar x-i\epsilon], \nonumber \\
  g_{V_2} &=& \left( {4\over 3}\ln \frac{m_b}{\mu}+{2\over 3}\right)\int^1_0\Phi^{V_2,h}(x)dx, \nonumber \\
   g'_{V_2} &=& {4\over 3}\ln \frac{m_b}{\mu}\int^1_0\Phi^{V_2,h}(x)dx,
 \end{eqnarray}
with $\Phi^{V_2,0}=\Phi^{V_2}_\|$, $\Phi^{V_2,\pm}=\Phi^{V_2}_\pm$.
For $i=6$, the result for the penguin contribution is
\begin{eqnarray} \label{eq:P6}
   P_6^{h,p}(V_2)&=&\frac{C_F\alpha_s}{4\pi N_c}\,\Bigg\{ \!
    c_1 \hat G^h_{V_2}(s_p)
    + c_3\bigg[ \hat G^h_{V_2}(s_s) + \hat G^h_{V_2}(1)
    \bigg]\nonumber \\
    &&+ (c_4+c_6)\sum_{i=u}^b  \hat G^h_{V_2}(s_i) \! \Bigg\},
    \hspace{0.5cm}
\end{eqnarray}
where the function $\hat G_{V_2}(s)$ is
defined as
 \begin{eqnarray} \label{eq:P6.1}
 \hat G^0_{V_2}(s) &=&  4\int^1_0 du\, \Phi_{v_2}(u) \int^1_0 dx\,x\bar x
  \ln[s-\bar u x\bar x-i\epsilon], \nonumber \\
 \hat G^\pm_{V_2}(s) &=& 0\,.
 \end{eqnarray}

The transverse penguin contractions vanish for $i=6,8$:
$P_{6,8}^{\pm,p}=0$. The $r_\chi^{M_2}$ term in
Eq. (\ref{eq:P6}) is factorized out so that when the vertex correction $V_{6,8}$ is neglected, $a_6^0$
contributes to the decay amplitude in the product $r_\chi^{V_2}a_6^0\approx r_\chi^{V_2}P_6^0$ {\cite{qcdf1,qcdf1-1}}.
For $i=8,10$,
\begin{equation}
   P_8^{h,p}(V_2) =  \frac{\alpha_{\rm em}}{9\pi N_c}\,(c_1+N_c c_2)\,
   \hat G^h_{V_2}(s_p) \,,
\end{equation}
\begin{equation}\label{PKEW}
   P_{10}^{h,p}(V_2) = \frac{\alpha_{\rm em}}{9\pi N_c} \{
   (c_1+N_c c_2)[ G^h_{V_2}(s_p)+2g_{V_2}]
   - 3c^{\rm eff}_\gamma  G_g^h \}.
\end{equation}
For $i=7,9$,
 \begin{eqnarray}\label{radphoton}
  P_{7,9}^{-,p}(V_2)
 &=& -{\alpha_{\rm em}\over 3\pi}C_{7\gamma}^{\rm eff}{m_Bm_b\over m_{V_2}^2}
 +{2\alpha_{\rm em}\over 27\pi}(c_1+N_c c_2)   \nonumber \\
  && \left[\delta_{pc}\ln{m_c^2\over\mu^2}+\delta_{pu}\ln{\nu^2\over \mu^2}+1\right].
 \end{eqnarray}

The relevant integrals for the dipole operators $O_{g,\gamma}$ are {\cite{qcdf1,qcdf1-1}}
  \begin{eqnarray}
&&  G^0_g = \int^1_0 du\,{\Phi^{V_2}_\|(u)\over \bar u}\,, \nonumber\\
&&  G^\pm_g =0.
 \label{eq:cg}
 \end{eqnarray}
The dipole
operators $Q_{8g}$ and $Q_{7\gamma}$ do not contribute to the
transverse penguin amplitudes at ${\cal O}(\alpha_s)$ due to angular
momentum conservation {\cite{Kagan2004}}.

\vskip 0.2in \noindent{\it \underline{Annihilation contributions}} \vskip 0.1in

 The annihilation contributions to the decay  $\overline B\to
V_{1}V_2$ can be described in terms of $b_i^{p,h}$ and $b_{i,{\rm EW}}^{p,h}$
\begin{eqnarray}\label{eq:h1ksann}
&&\frac{G_F}{\sqrt2} \sum_{p=u,c} \! \lambda_p\, \!\langle V_{1}V_2
|{\cal T_B}^{h,p} |\overline B^0\rangle        \nonumber \\
&=&i\frac{G_F}{\sqrt{2}}\sum_{p=u,c} \lambda_p
 f_B f_{V_1} f_{V_{2}}\sum_i (b_i^{p,h}+b_{i,{\rm EW}}^{p,h}).
\end{eqnarray}
The building blocks have the expressions
 \begin{eqnarray} \label{eq:bi}
 b_1 &=& {C_F\over N_c^2}c_1A_1^i, \qquad\quad b_2 = {C_F\over N_c^2}c_2A_1^i, \nonumber \\
 b_3&=&{C_F\over N_c^2}\left[c_3A_1^i+c_5(A_3^i+A_3^f)+N_cc_6A_3^f\right], \nonumber \\
  b_4&=&{C_F\over N_c^2}\left[c_4A_1^i+c_6A_2^f\right], \nonumber \\
 b_{\rm 3,EW} &=& {C_F\over
 N_c^2}\left[c_9A_1^{i}+c_7(A_3^{i}+A_3^{f})+N_cc_8A_3^{i}\right],
 \nonumber \\
 b_{\rm 4,EW} &=& {C_F\over
 N_c^2}\left[c_{10}A_1^{i}+c_8A_2^{i}\right],
 \end{eqnarray}
where we have omitted the superscripts $p$ and $h$ in above
expressions for simplicity. The subscripts 1,2,3 of $A_n^{i,f}$ denote the annihilation
amplitudes induced from $(V-A)(V-A)$, $(V-A)(V+A)$ and $(S-P)(S+P)$ operators,
respectively, and the superscripts $i$ and $f$ refer to gluon emission from the
initial and final-state quarks, respectively. $V_1$  contains an antiquark from the weak vertex and $V_2$
contains a quark from the weak vertex {\cite{qcdf1,qcdf1-1}}. The explicit expressions of weak
annihilation amplitudes are:
\begin{eqnarray}
A_1^{i,\,0}(V_{1} V_{2}) &=&
 \pi\alpha_s \int_0^1\! du \,dv\, \nonumber \\
&& \Bigg\{\Phi_\parallel^{V_1}(v) \Phi_\parallel^{V_2}(v)\,
\left[\frac{1}{u(1-\bar u v)}+\frac{1}{u\bar v^2}\right]
 \nonumber\\
 &-& r_\chi^{V_1}r_\chi^{V_2} \,\Phi_{v_1}(u)\, \Phi_{v_2}(v)
\frac{2}{u\bar{v}} \Bigg\}\,, \label{eq:A1i0} \\ \nonumber
 A_1^{i,\,-}(V_1 V_2)  &=& -
 \pi\alpha_s {2m_{V_1}m_{V_2}\over m_B^2}\int_0^1\! du \,dv\, \nonumber \\
&& \Bigg\{
 \,\Phi^{V_1}_-(u)\,\Phi_-^{V_2} (v)\Bigg[  \frac{\bar u+\bar v}{u^2 \bar{v}^2}\nonumber \\
&& +{1\over (1-\bar u v)^2}\Bigg]
 \Bigg\}, \\ \nonumber
  A_1^{i,\,+}(V_1 V_2)  &=& -
 \pi\alpha_s {2m_{V_1}m_{V_2}\over m_B^2}\int_0^1\! du \,dv\,  \nonumber \\
 &&\Bigg\{ \,\Phi^{V_1}_+(u)\,\Phi_+^{V_2} (v)\Bigg[ {2\over u\bar v^3}
 - \frac{v}{(1-\bar uv)^2}\nonumber\\
 &&-{v\over \bar v^2(1-\bar u v)}\Bigg]
 \Bigg\}\,,\hspace{1cm}\label{eq:A1ip}\\ \nonumber
A_2^{i,\,0}(V_{1} V_{2}) &=&
 \pi\alpha_s \int_0^1\! du \,dv\, \Bigg\{
 \Phi_\parallel^{V_1}(v) \Phi_\parallel^{V_2}(v)\, \nonumber \\
&& \Bigg[\frac{1}{\bar v(1-\bar u v)}
 +\frac{1}{u^2\bar v}\Bigg] \nonumber \\
  &&-r_\chi^{V_1}r_\chi^{V_2} \,\Phi_{m_1}(u)\, \Phi_{m_2}(v)
\frac{2}{u\bar{v}} \Bigg\}\,, \label{eq:A2i0} \\ \nonumber
 A_2^{i,\,-}(V_1 V_2)  &=& -
 \pi\alpha_s {2m_{V_1}m_{V_2}\over m_B^2}\int_0^1\! du \,dv\,\nonumber \\
 &&\Bigg\{ \Phi^{V_1}_+(u)\,\Phi_+^{V_2} (v)
 \nonumber \\
 && \times \left[\frac{u+v}{u^2 \bar{v}^2}
 +{1\over (1-\bar u v)^2}\right]
 \Bigg\}\,,  \\  \nonumber
   A_2^{i,\,+}(V_1 V_2)  &=& -
 \pi\alpha_s {2m_{V_1}m_{V_2}\over m_B^2}\int_0^1\! du \,dv\,\nonumber \\
 &&\Bigg\{\Phi^{V_1}_-(u)\,\Phi_-^{V_2} (v)
 \nonumber \\
 && \times\Bigg[ {2\over u^3\bar v}
   - \frac{\bar u}{(1-\bar uv)^2}\nonumber \\
 &&-{\bar u\over u^2(1-\bar u v)}\Bigg]
 \Bigg\}\,, \hspace{1cm}\label{eq:A2ip}  \nonumber \\
 \end{eqnarray}
  \begin{eqnarray}
A_3^{i,\,0}(V_{1} V_{2}) &=&
 \pi\alpha_s \int_0^1\! du \,dv\, \Bigg\{
 r_\chi^{V_1}\Phi_{m_1}(v) \Phi_\parallel^{V_2}(v)\,\nonumber \\
 &&\frac{2\bar u}{u\bar v(1-\bar u v)}
+r_\chi^{V_2} \,\Phi_\parallel^{V_1}(v)\, \Phi_{m_2}(v)\nonumber \\
&&\frac{2v}{u\bar{v}(1-\bar uv)} \Bigg\}, \label{eq:A3i0} \\ \nonumber
 A_3^{i,\,-}(V_1 V_2)  &=& -
 \pi\alpha_s \int_0^1\! du \,dv\,
 \Bigg\{ -{m_{V_2}\over m_{V_1}} r_\chi^{V_1}
 \,\Phi^{V_1}_\perp(u)\,   \nonumber \\
 &&\Phi_-^{V_2} (v)\frac{2}{u\bar v(1-\bar uv)}
 + {m_{V_1}\over m_{V_2}} r_\chi^{V_2}\nonumber \\
 &&\,\Phi^{V_1}_+(u)\,\Phi_\perp^{V_2} (v)\frac{2}{u\bar v(1-\bar
 uv)} \Bigg\}\,, \hspace{1cm}\label{eq:A3im}\\
A_3^{f,\,0}(V_{1} V_{2}) &=&
 \pi\alpha_s \int_0^1\! du \,dv\, \Bigg\{
  r_\chi^{V_1} \,\Phi_{m_1}(u) \Phi_\parallel^{V_2}(v)\,
 \frac{2(1+\bar v)}{u\bar{v}^2}
 \nonumber\\
 && - r_\chi^{V_2} \,\Phi_\parallel^{V_1}(u)\, \Phi_{m_2}(v)
\frac{2(1+u)}{u^2\bar{v}} \Bigg\}\,, \label{eq:A3f0} \\
 A_3^{f,\,-}(V_1 V_2)  &=& -
 \pi\alpha_s \int_0^1\! du \,dv\, \Bigg\{
  \frac{m_{V_2}}{m_{V_1}} r_\chi^{V_1} \,\Phi^{V_1}_{\perp}(u)\,\Phi_-^{V_2} (v)
 \frac{2}{u^2 \bar{v}}
 \nonumber\\
 &&+\frac{m_{V_1}}{m_{V_2}} r_\chi^{V_2}\,\Phi_+^{V_1}(u)\, \Phi^{V_2}_{\perp}(v) \,
 \frac{2}{u\bar v^2 }
 \Bigg\}\,, \label{eq:A3fm}
 \end{eqnarray}
and $A_1^{f,h}=A_2^{f,h}=A_3^{i,+}=A_3^{f,+}=0$.
$V_1$ contains an antiquark from the weak vertex with longitudinal fraction
$\bar y$, while $V_2$ contains a quark from the weak vertex with momentum
fraction $x$ \cite{qcdf1,qcdf1-1}.

$A_{1,2}^{i,\pm}$ are
suppressed by a factor of $m_1m_2/m_B^2$ relative to other terms, so
only the annihilation
contributions due to $A_3^{f,0}$, $A_3^{f,-}$, $A_{1,2,3}^{i,0}$ and $A_3^{i,-}$ are considered.

The logarithmic divergences in annihilation can be extract into unknown variable $X_A$
\begin{eqnarray}
 \int_0^1 {du\over u}\to X_A, \qquad \int_0^1 {\ln u\over u}\to -{1\over 2}X_A.
 \end{eqnarray}

\subsection{\label{sec:CP form}The calculation of CP violation}
In order to obtain the $CP$ violation of $\bar{B}\rightarrow \rho^{0}(\omega)\rho^{0}(\omega)\rightarrow \pi^{+}\pi^{-}\pi^{+}\pi^{-}$ in Eq.{(\ref{eq:CP-tuidao})},
we calculate the amplitudes $t_\rho$, $t_\rho^a$, $t_\omega$, $t_\omega^a$,
$p_{\rho}$, $p_{\rho}^{a}$, $p_{\omega}$ and $p_{\omega}^{a}$ in Eqs.{(\ref{tree-scena1})}{(\ref{pengui-scena1})} in the QCDF approach, which are
tree-level and penguin-level amplitudes.
The decay amplitudes for the process $\bar{B}\rightarrow \rho^{0}\rho^{0}(\omega)$ are in the QCD factorization as follows:
\begin{eqnarray}
A_{\bar{B}\rightarrow \rho^{0}\rho^{0}}&=&
A_{\rho^{0}\rho^{0}}(\alpha_{4}^{p}-\delta_{pu}\alpha_2-\frac{1}{2}\alpha_{4,EW}^{p}
-\frac{3}{2}\alpha_{3,EW}^{p}  \nonumber \\
&+&\beta_3^{p}-\frac{1}{2}\beta_{3,EW}^{p}
+\delta_{pu}\beta_1+2\beta_{4}^{p}+\frac{1}{2}\beta_{4,EW}^{p}),  \nonumber \\
\end{eqnarray}
\begin{eqnarray}
-2A_{\bar{B}\rightarrow \rho^{0}\omega}&=&
A_{\rho^{0}\omega}(\delta_{pu}\alpha_2-\delta_{pu}\beta_1+2\alpha_{3}^{p}+\alpha_{4}^{p} \nonumber \\
&+&\frac{1}{2}\alpha_{3,EW}^{p}-\frac{1}{2}\alpha_{4,EW}^{p}+\beta_3^{p}-\frac{1}{2}\beta_{3,EW}^{p} \nonumber \\
&-&\frac{3}{2}\beta_{4,EW}^{p})+A_{\omega\rho^{0}}(-\delta_{pu}\alpha_2-\delta_{pu}\beta_1+\alpha_{4}^{p}  \nonumber \\
&-&\frac{3}{2}\alpha_{3,EW}^{p}-\frac{1}{2}\alpha_{4,EW}^{p}+\beta_3^{p}-\frac{1}{2}\beta_{3,EW}^{p}\nonumber \\
&-&\frac{3}{2}\beta_{4,EW}^{p}),
\end{eqnarray}
where
\begin{eqnarray}
A_{V_{1}V_{2}}=i\frac{G_{F}}{\sqrt{2}}\langle V_1|(\bar{q}b)_{V-A}|B\rangle
\langle V_2|(\bar{q}q)_{V}|0\rangle
\end{eqnarray}

From Eq.{(\ref{tree-form})}, one can get
\begin{eqnarray}
\alpha e^{i\delta_{\alpha}}=\frac{t_\omega+t_\omega^a}{t_\rho+t_\rho^a}
=\frac{Q_2}{Q_1},
\end{eqnarray}
where
\begin{eqnarray}
Q_1&=&t_{\rho}+t_{\rho}^{a}        \nonumber \\
&=&A_{\rho^{0}\rho^{0}}[\alpha_4^{u,h}-\alpha_4^{c,h}-\delta_{pu}\alpha_2   \nonumber \\
&-&\frac{1}{2}(\alpha_{4,EW}^{u,h}
-\alpha_{4,EW}^{c,h})+\delta_{pu}\beta_1]\\
Q_2&=&t_{\omega}+t_{\omega}^{a}      \nonumber \\
   &=&-\frac{1}{2}A_{\rho^{0}\omega^{0}}[\delta_{pu}\alpha_2-\delta_{pu}\beta_1
   +\alpha_4^{u,h}   \nonumber \\
   &&-\alpha_4^{c,h}-\frac{1}{2}(\alpha_{4,EW}^{u,h}
-\alpha_{4,EW}^{c,h}]\nonumber \\
&&-\frac{1}{2}A_{\omega^{0}\rho^{0}}[-\delta_{pu}\alpha_2-\delta_{pu}\beta_1
+\alpha_4^{u,h}\nonumber \\
&&-\alpha_4^{c,h}-\frac{1}{2}(\alpha_{4,EW}^{u,h}
-\alpha_{4,EW}^{c,h}]
\end{eqnarray}

In a similar way, with the aid of the Fierz identities, we can
evaluate the penguin operator contributions $p_{\rho}$ and
$p_{\omega}$. From Eq. {(\ref{p-form})} we have
\begin{eqnarray}
\beta e^{i\delta_{\beta}}=\frac{p_{\rho}+p_{\rho}^{a}}{p_{\omega}+p_{\omega}^{a}}
=\frac{Q_3}{Q_4},
\end{eqnarray}
where
\begin{eqnarray}
Q_3&=&p_{\rho}+p_{\rho}^{a}       \nonumber \\
&=&A_{\rho^{0}\rho^{0}}[(-\frac{1}{2})\alpha_{4,EW}^{c,h}-(\frac{3}{2})\alpha_{3,EW}^{c,h}] \nonumber \\
&&+\beta_3^{p}-\frac{1}{2}\beta_{3,EW}^{p}+2\beta_{4}^{p}+\frac{1}{2}\beta_{4,EW}^{p})\\
Q_4&=&p_{\omega}+p_{\omega}^{a}     \nonumber \\
&=&-\frac{1}{2}A_{\rho^{0}\omega}[2\alpha_{3}^{c}+\alpha_{4}^{c}]+\frac{1}{2}\alpha_{3,EW}^{c}\nonumber \\
&&-\frac{1}{2}\alpha_{4,EW}^{c}
+\beta_3^{p}-\frac{1}{2}\beta_{3,EW}^{p}-\frac{3}{2}\beta_{4,EW}^{p})   \nonumber \\
&&-\frac{1}{2}A_{\omega\rho^{0}}(\alpha_{4}^{c}
-\frac{3}{2}\alpha_{3,EW}^{c}-\frac{1}{2}\alpha_{4,EW}^{c}\nonumber \\
&&+\beta_3^{p}-\frac{1}{2}\beta_{3,EW}^{p}
-\frac{3}{2}\beta_{4,EW}^{p})
\end{eqnarray}

Form Eq. {(\ref{p2-form})} we have
\begin{eqnarray}
r'e^{i(\delta_{q}+\phi)}=\frac{p_{\omega}+p_{\omega}^{a}}{t_{\rho}+t_{\rho}^{a}}
=\frac{Q_4}{Q_1},
\end{eqnarray}

\begin{equation}
r'e^{i\delta_{q}}=\frac{Q_4}{Q_1}
\left|\frac{V_{tb}V_{td}^{*}}{V_{ub}V_{ud}^{*}}\right|,
\end{equation}
where
\begin{eqnarray}
\left|\frac{V_{tb}V_{td}^{*}}{V_{ub}V_{ud}^{*}}\right|=
\frac{\sqrt{(1-\rho)^{2}+\eta^{2}}}{(1-\frac{\lambda^{2}}{2})(\sqrt{\rho^{2}+\eta^{2}})}.
\end{eqnarray}

\section{\label{input}Input parameters}
\label{sec:7}In the numerical calculations, we should input
distribution amplitudes and the CKM matrix elements in the
Wolfenstein parametrization. For the CKM matrix elements, which are
determined from experiments, we use the results in Ref.
\cite{J2012}:
\begin{eqnarray}
\bar{\rho}=0.132^{+0.022}_{-0.014}, &&\hspace{1cm}  \bar{\eta}=0.341\pm 0.013,  \nonumber \\
\lambda=0.2253\pm0.0007, &&\hspace{1cm}  A=0.808^{+0.022}_{-0.015},
\end{eqnarray}
 where
\begin{eqnarray}
 \bar{\rho}=\rho(1-\frac{\lambda^2}{2}),\quad
\bar{\eta}=\eta(1-\frac{\lambda^2}{2}).\label{eq: rho rhobar
relation}
\end{eqnarray}

The general expressions of the helicity-dependent amplitudes
can be simplified by considering the asymptotic distribution amplitudes
for $\Phi_V,\Phi_v$ :
\begin{eqnarray}
 \Phi^V_\parallel(u)=6u\bar u, &&\quad   \Phi_v(u)=3(2u-1),       \nonumber \\
 \Phi_\perp^V(u)=6u\bar u, &&\quad   \Phi^V_+=\int^1_u dv{\Phi^V_\parallel(v)\over v},    \nonumber \\
\Phi^M_-&=&\int^u_0 dv{\Phi^V_\parallel(v)\over \bar v}.
\end{eqnarray}

Power corrections in QCDF always involve endpoint divergences which produce some uncertainties. The endpoint
divergence $X\equiv\int^1_0 dx/\bar x$ in the annihilation and hard spectator
scattering diagrams is parameterized as
\begin{eqnarray}\label{eq:XA}
 X_A&=&\ln\left({m_B\over \Lambda_h}\right)(1+\rho_A e^{i\phi_A}), \qquad \\
 X_H&=&\ln\left({m_B\over \Lambda_h}\right)(1+\rho_H e^{i\phi_H}),
\end{eqnarray}
with the unknown real parameters $\rho_{A,H}$ and $\phi_{A,H}$ {\cite{qcdf1,qcdf1-1}}. For simplicity,
we shall assume that $X_A^h$ and $X_H^h$ are helicity independent:
$X_A^-=X_A^+=X_A^0$ and $X_H^-=X_H^+=X_H^0$.

\section{\label{sec:numerical}Numerical results}

\begin{figure}
% Use the relevant command for your figure-insertion program
% to insert the figure file.
% For example, with the option graphics use
\resizebox{0.5\textwidth}{!}{%
  \includegraphics{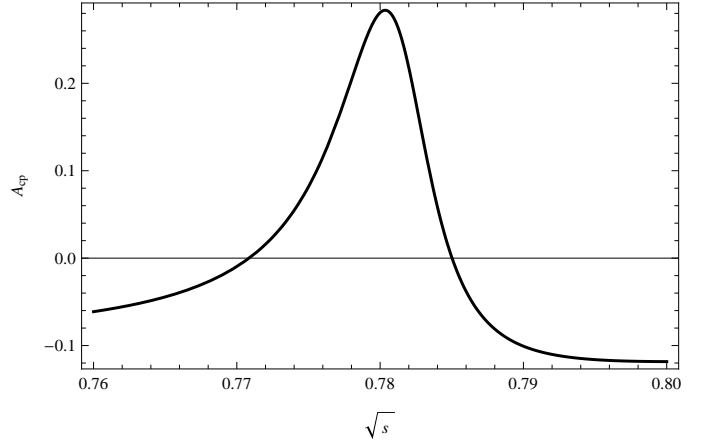}
  }
% If not, use
%\vspace{5cm}       % Give the correct figure height in cm
\caption{Plot of $A_{CP}$ as a function of $\sqrt{s}$
corresponding to central parameter values of CKM matrix elements
for $\bar{B}^{0}\rightarrow \rho^{0}(\omega)\rho^{0}(\omega)\rightarrow \pi^{+}\pi^{-}\pi^{+}\pi^{-}$.
}
\label{fig:1}       % Give a unique label
\end{figure}

\begin{figure}
% Use the relevant command for your figure-insertion program
% to insert the figure file.
% For example, with the option graphics use
\resizebox{0.5\textwidth}{!}{%
  \includegraphics{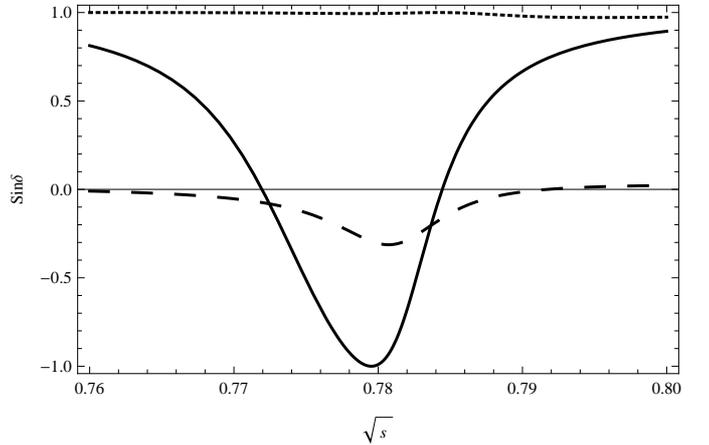}
  }
% If not, use
%\vspace{5cm}       % Give the correct figure height in cm
\caption{Plot of $\sin\delta$ as a function of $\sqrt{s}$
corresponding to central parameter values of CKM matrix elements
for $\bar{B}^{0}\rightarrow \rho^{0}(\omega)\rho^{0}(\omega)\rightarrow \pi^{+}\pi^{-}\pi^{+}\pi^{-}$.
The solid (dashed and dotted) line corresponds to $\sin\delta_{0}$ ($\sin\delta_{-}$ and $\sin\delta_{+}$) respectively.}
\label{fig:1}       % Give a unique label
\end{figure}

\begin{figure}
% Use the relevant command for your figure-insertion program
% to insert the figure file.
% For example, with the option graphics use
\resizebox{0.5\textwidth}{!}{%
  \includegraphics{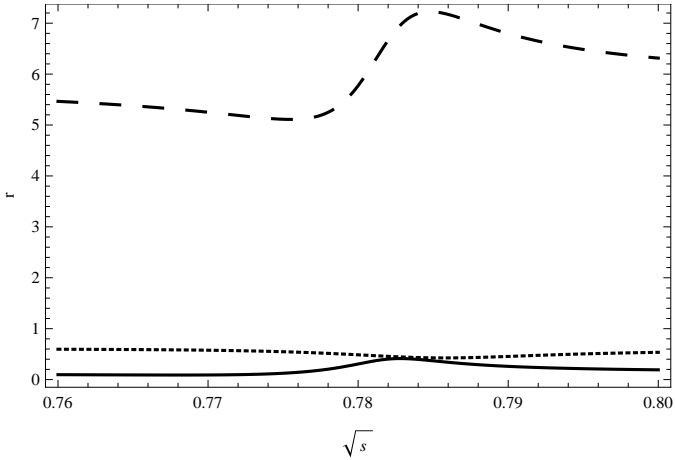}
  }
% If not, use
%\vspace{5cm}       % Give the correct figure height in cm
\caption{Plot of $r$ as a function of $\sqrt{s}$
corresponding to central parameter values of CKM matrix elements
for $\bar{B}^{0}\rightarrow \rho^{0}(\omega)\rho^{0}(\omega)\rightarrow \pi^{+}\pi^{-}\pi^{+}\pi^{-}$.
The solid (dashed and dotted) line corresponds to $r_{0}$ ($r_{-}$ and $r_{+}$) respectively.}
\label{fig:1}       % Give a unique label
\end{figure}

In the numerical results, we find that for the decay channel we are considering
the $CP$ violation can be enhanced via  $\rho-\omega$ mixing when the
invariant mass of $\pi^{+}\pi^{-}$ is in the vicinity of the
$\omega$ resonance. The uncertainties of the CKM matrix elements mainly come from $\rho$ and $\eta$. In our numerical
results, we let $\rho$ and $\eta$ vary between the limiting values.
We find the results are not sensitive to the  values of $\rho$ and $\eta$.
Hence, the numerical results are shown in Fig.1, Fig.2 and Fig.3 with the central parameter values of
CKM matrix elements. From the numerical results, it is found that
there is a maximum $CP$ violating parameter value, $A_{CP}^{max}$,
when the masses of the $\pi^+\pi^-$ pairs are in the vicinity of the $\omega$ resonance.
In Fig.1, one can find that the maximum $CP$ violating parameter reaches
$28\%$ in the case of ($\rho_{central}$, $\eta_{central}$).

From the Eq.(\ref{eq:CP-tuidao}) one can find that the
$CP$ violating parameter is related to
$\sin\delta$ and $r$.
In Fig.2, we show the plot of $\sin\delta_{0}$ ($\sin\delta_{-}$ and $\sin\delta_{+}$)
as a function of $\sqrt{s}$.
We can see that the $\rho-\omega$ mixing
mechanism produces a large $\sin\delta_{0}$ ($\sin\delta_{-}$ and $\sin\delta_{+}$)
at the $\omega$ resonance.
As can be seen from Fig.2, the plots vary sharply in the cases of
 $\sin\delta_{0}$ and $\sin\delta_{-}$. Meanwhile, $\sin\delta_{+}$ changes weakly comparing with
the $\sin\delta_{0}$ and $\sin\delta_{-}$.
It can be seen from the Fig.3 that $r_0$ and $r_{-}$ change more rapidly than $r_{+}$
when the masses of the $\pi^+\pi^-$ pairs are in the vicinity of the $\omega$ resonance.

In the paper {\cite{gang3}}, we studied the enhanced $CP$ violation for the decay channel
$\bar{B}^{0}\rightarrow \pi^{+}\pi^{-}\pi^{+}\pi^{-}$ in the naive factorization.
Since non-factorizable contribution can not be calculated in the naive factorization, $N_c$ was treated as an effective
parameter. We found that the $CP$ violating asymmetry
was large and ranges from $-82\%$ to $-98\%$ via the $\rho-\omega$ mixing mechanism strongly depending on the value $N_c$
when the invariant mass of the $\pi^{+}\pi^{-}$ pair is in
the vicinity of the $\omega$ resonance. However, the maximum $CP$ violation only can reach
$28\%$ via double $\rho-\omega$ mixing in the QCD factorization.
The naive factorization scheme has been shown to be the leading order result in the framework of
QCD factorization when the radiative QCD corrections $O(\alpha_{s}(m_b))$  and the order
$O(1/m_b)$ effects are neglected. The QCD factorization can evaluate systematically corrections to the
results from the naive factorization. The distinction between the naive factorization and the QCDF mainly come from
the strong phases of the QCD corrections. In the calculating process, we find that the annihilation contributions in QCDF
which introduce the unknown parameters are small. Hence, the uncertainties of the results from the QCDF become small.

\section{\label{sec:conclusion}Summary and conclusions}

In this paper, we studied the $CP$ violation for the decay process
$\bar{B}^{0}\rightarrow \pi^{+}\pi^{-}\pi^{+}\pi^{-}$
due to the interference of $\rho-\omega$ mixing
in the QCDF approach. This process induces two
$\rho-\omega$ interference. It was found
the $CP$ violation can be enhanced at
the region of $\rho-\omega$ resonance.
As a result, the maximum $CP$ violation
could reach $28\%$. $\rho-\omega$ mixing is small due to the isospin violation.  However, the mixing can produce
a large strong phase, $\delta$, in Eq. (\ref{jianhua1}). This is because when the invariant masses of the $\pi^{+}\pi^{-}$ pairs are in the vicinity of $\omega$, $s_{\omega}\sim im_{\omega}\Gamma_{\omega}$,
and it becomes comparable with $\tilde{\Pi}_{\rho\omega}$ in Eq. (\ref{jianhua1}). In other words, $\rho-\omega$ mixing becomes important in the vicinity of $\omega$.
This is the reason why we can see large CP violation in the vicinity of $\omega$.
Beyond the $\rho-\omega$ interference region, the noticeable values of CP violation are caused  by the strong phases provided by the Wilson coefficients.

The LHC experiments are designed with the
center-of-mass energy $14$ TeV and the luminosity $L=10^{34}
cm^{-2}s^{-1}$. The heavy quark physics is one of
the main topics of LHC experiments.
Especially, LHCb detector is designed to
make precise studies on $CP$ asymmetries
and rare decays of b-hadron systems.
Recently, the LHCb Collaboration found clear evidence for
direct $CP$ violation in some three-body decay channels in
charmless decays of $B$ meson.
Large $CP$ violation is observed in
$B^{+}\rightarrow K^{+}K^{-}\pi^{+}$,
$B^{\pm}\rightarrow \pi^{\pm}\pi^{+}\pi^{-}$ in the region
$m^{2}_{\pi^{+}\pi^{-}low}<0.4$ GeV${^2}$ and $m^{2}_{\pi^{+}\pi^{-}low}>15$ GeV${^2}${\cite{J-R.A}}.
LHCb experiment may collect data in the region of the invariant masses of $\pi^{+}\pi^{-}$
associated the $\omega$ resonance for detecting our prediction of $CP$ violation.

In our calculations there are some uncertainties.
The QCD factorization scheme provides a framework in
which we can evaluate systematically corrections to the results
obtained in the naive factorization scheme. However, when we take
into account the nonfactorizable and chirally enhanced
hard-scattering spectator and annihilation contributions which
appear at order $O(\alpha_{s}(m_b))$ and  $O(1/m_b$), respectively,
the involvement of the twist-3 hadronic distribution amplitudes
leads to logarithmical divergence coming from the endpoint
integrals. This brings large uncertainties in the predictions of the
$CP$ violating asymmetries in the QCD factorization scheme.
Furthermore, in addition to the model dependence appearing in the
factorized hadronic matrix elements just as in the naive
factorization scheme, we cannot avoid the model dependence and
process dependence of the hard-scattering spectator and annihilation
contributions due to their dependence on the hadronic distribution
amplitudes and dependence on different processes. Such dependence
will also appear if one tries to include other $1/m_b$ corrections
and even higher order corrections. This leads to uncertain of
our results.

\paragraph{Acknowledgments}
This work was supported by National Natural Science
Foundation of China (Project Numbers 11147003, 11175020, 11275025, 11347124
and U1204115), Plan For Scientific Innovation Talent of Henan University of Technology
 (Project Number 2012CXRC17), the Key Project (Project Number 14A140001)
 for Science and Technology of the Education Department Henan Province.

%\newpage

%\newpage

\end{document}